\documentclass[a4paper,fleqn]{cas-sc}

\usepackage[numbers]{natbib}
\def\tsc#1{\csdef{#1}{\textsc{\lowercase{#1}}\xspace}}
\tsc{WGM}
\tsc{QE}
\tsc{EP}
\tsc{PMS}
\tsc{BEC}
\tsc{DE}

 \usepackage{amsmath,amsthm,amssymb,amsfonts}
\usepackage{array}

\usepackage{graphicx}
\usepackage{textcomp}
\usepackage[mathscr]{eucal}
\usepackage[ruled,vlined,linesnumbered]{algorithm2e}
\usepackage{algpseudocode}
\usepackage{multirow}
\usepackage{rotating}

\newtheorem{example}{Example}
\usepackage[font=small]{caption}
\usepackage{subcaption}
\usepackage{pifont}
\usepackage{times}
\usepackage{fancyhdr, graphicx, amsmath,amssymb}
\algnewcommand\LeftComment[2]{%
\hspace{#1\algindent}$\triangleright$ \eqparbox{COMMENT}{#2} \hfill %
}
\newcommand{\nonl}{\renewcommand{\nl}{\let\nl\oldnl}}

\begin{document}
\let\WriteBookmarks\relax
\def\floatpagepagefraction{1}
\def\textpagefraction{.001}

\shorttitle{Optimal Task Allocation and Resource Aware Containerization}

\shortauthors{Kumar and Mishra}

\title [mode = title]{An Auction-Based Mechanism for Optimal Task Allocation and Resource Aware Containerization}                      

\author[1]{Ramakant Kumar}
\ead{ramakant.kumar@galgotiacollege.edu}


\affiliation[1]{organization={Department of Data Science},
    addressline={GALGOTIAS COLLEGE OF ENGINEERING AND TECHNOLOGY}, 
    city={Noida},
    country={India}}

\cortext[cor1]{Corresponding author}

\begin{abstract}
 Distributed computing has enabled cooperation between multiple computing devices for the simultaneous execution of resource-hungry tasks. Such execution also plays a pivotal role in the parallel execution of numerous tasks in the Internet of Things (IoT) environment. Leveraging the computing resources of multiple devices, the offloading and processing of computation-intensive tasks can be carried out more efficiently. However, managing resources and optimizing costs remain challenging for successfully executing tasks in cloud-based containerization for IoT. This paper proposes AUC-RAC, an auction-based mechanism for efficient offloading of computation tasks among multiple local servers in the context of IoT devices. The approach leverages the concept of Docker swarm, which connects multiple local servers in the form of Manager Node (MN) and Worker Nodes (WNs). It uses Docker containerization to execute tasks simultaneously. In this system, IoT devices send tasks to the MN, which then sends the task details to all its WNs to participate in the auction-based bidding process. The auction-based bidding process optimizes the allocation of computation tasks among multiple systems, considering their resource sufficiency. The experimental analysis establishes that the approach offers improved offloading and computation-intensive services for IoT devices by enabling cooperation between local servers.
\end{abstract}

\begin{highlights}
\item Distributed computing fosters collaboration for efficient execution of resource-hungry IoT tasks.  
\item AUC-RAC uses auction-based bidding for task offloading via Docker swarm in IoT environments.  
\item Experimental results highlight AUC-RAC's ability tooptimisee task allocation and computation services.  
\end{highlights}


\maketitle

\section{Funding}The authors declare that no funds, grants, or other support were received during the preparation of this manuscript.
\section{Introduction}
Internet of Things (IoT) has grown exponentially in every field, like healthcare, banking, manufacturing, consumer goods, and so on.  Traditionally, IoT devices are connected to the physical infrastructure for storing and processing the data generated by the sensor deployed in the environment. The physical infrastructure comprises servers, where the operating system runs over the hardware. The hardware includes memory, network cards, and a microcontroller chip. The operating systems (such as Windows, Linux, etc.) can be installed on the hardware depending upon their application demands. In IoT, there is a need for resource scaling to meet dynamic changes. Physical infrastructure: we can not facilitate resource scaling to bear out dynamic changes\cite{naha2021multi}. In traditional physical infrastructure, the platform dependency of built applications limits widespread acceptability.

To address the aforementioned shortcomings, the industry has adopted virtualization, enabling the creation of multiple independent Virtual Machines (VMs) on physical infrastructure via hypervisors. Each VM operates with its own operating system, allowing it to function independently while utilizing a shared pool of hardware resources~\cite{rrr4, henkel2020learning, rrr10, rrr12}. This setup optimizes resource utilization through resource adjustment and sharing. However, VMs are not ideal for low-processing and resource-constrained devices due to the overhead of requiring an operating system for each VM. This limitation makes them less suitable for IoT environments, where efficient and lightweight solutions like Docker containerization are more appropriate for executing tasks on devices.

To address the challenges of efficient task execution and resource management, industries have increasingly adopted containerization techniques~\cite{anderson2015Docker, rrr5, bernstein2014containers,rrr11}. In recent years, Docker containerization has gained popularity among software development companies due to its lightweight nature and efficiency~\cite{bhimani2018Docker,rrr2, manu2016study, qiao2016doopnet, rrr1}. Unlike traditional virtual machines, Docker containers share the host machine's operating system, eliminating the need to install a separate OS for each container. This results in faster task execution and efficient resource utilization. Docker Engine, special software installed on the host OS, manages the creation and operation of Docker containers. Containers can pull required software from the Docker Hub as needed, and Docker images can be shared and reused across different environments, supporting a platform-as-a-service (PaaS) model~\cite{smet2018Docker, singh2016containers}. Docker's layered image structure allows for lightweight and rapid image transfer. The concept of Docker Swarm facilitates the integration of multiple local servers, allowing them to operate as Manager Nodes (MNs) and Worker Nodes (WNs) for executing offloaded IoT tasks~\cite{cito2017empirical, rrr6, rrr9, rrr13}. Each node in the swarm can host one or more containers, enabling a robust and scalable distributed computing environment.

Effectively managing resources and optimizing costs in cloud-based containerization for IoT environments remain significant challenges. Traditional methods often struggle to allocate computation tasks efficiently and ensure timely execution within budget constraints. This paper explores \textit{how IoT devices can offload tasks to MNs and how MNs, in turn, distribute these tasks among WNs.} Given that MNs and WNs function as servers, \textit{how can Docker containerization be utilized to execute a variety of platform-dependent tasks concurrently?} This paper addresses these challenges by proposing a novel mechanism that optimizes resource allocation and minimizes costs, enhancing the efficiency of IoT-driven task execution. This paper introduces \textbf{AUC-RAC}, an innovative auction-based mechanism designed to optimize task allocation and resource management in IoT environments through containerization.

\begin{itemize}
\item \textit{IoT Device Task Offloading:} The paper proposes a method for offloading tasks from IoT devices to multiple local servers, enabling efficient execution. These servers are interconnected using the Docker swarm concept, forming a network of MNs and WNs.

\item \textit{Auction-Based Task Allocation:} AUC-RAC utilizes an auction-based bidding process to allocate tasks from the MN to WNs. This approach optimizes task distribution based on resource availability and cost, ensuring efficient and cost-effective allocation.

\item  \textit{Docker Containerization:} All WNs employ Docker containerization techniques, allowing them to execute many IoT tasks simultaneously, regardless of platform dependencies. This approach enhances resource utilization and reduces execution time.

\item  \textit{Profit and Cost Optimization:} The MN coordinates task allocation and optimizes its profit by effectively managing the costs paid to WNs. This profit optimization ensures sustainability and efficiency in task execution. AUC-RAC is unique in ensuring cost optimization at the task distribution and execution levels, making it a robust solution for IoT-driven environments where resource management and cost efficiency are critical.

\item \textit{Experimental Validation:} The paper provides a comprehensive experimental analysis demonstrating the efficacy of AUC-RAC. The results indicate significant improvements in offloading and executing computation-intensive tasks, offering enhanced service delivery for IoT devices.

\end{itemize}

\noindent \textbf{Roadmap:} The remaining part of the paper is as follows. The next section describes the background and motivation. Section~\ref{applications} highlights preliminaries and problem statements and  Section~\ref{sec5} presents our proposed AUC-RAC approach.  The experimental results and conclusion of this work are presented in Section~\ref{experiment} and Section~\ref{conclusion}, respectively.

\section{Background}\label{background}

\subsection{Interoperability in IoT Systems}  
A significant barrier to IoT adoption is the lack of interoperability among devices and platforms. Container-based virtualization has emerged as a promising solution, enabling dynamic application updates and efficient resource usage. Mekonnen \textit{et al.}~\cite{g4} analyzed energy efficiency in IoT environments, finding a $13\%$ power overhead in camera nodes using Docker during boot-up and shutdown. Similarly, the authors in~\cite{10.1145/3586010} \textbf{IoTvar}, a middle-ware library that simplifies IoT application development by automating sensor variable management and integration with platforms like FIWARE, OM2M, and muDEBS. It reduces developer effort with minimal impact on resource usage but limits advanced customizations.

\subsection{Virtualization in Real-Time Systems} 
Virtualization techniques have been extensively studied for real-time applications. Sollfrank \textit{et al.}~\cite{g1} investigated how Docker and other lightweight container technologies affect latency and response times in industrial automation. Their findings highlighted the role of effective resource management in achieving reliable performance. The work in~\cite{10406151} proposes a Supply Graph-based Partitioning (SGP) framework for efficient network virtualization in real-time wireless networks, addressing node dependency and timing constraints. SGP reduces computational overhead while achieving results comparable to an exact SMT-based solution. Similarly, the authors in~\cite{10144238} evaluate the performance of lightweight virtualization in industrial automation using the Siemens Industrial Edge framework, achieving round-trip latencies of $10$ ms, suitable for process control and supervision. It highlights the potential of edge computing to meet dynamic production needs in Industry 4.0.

\subsection{Service Handoff in Edge Computing}  
Service handoff remains a critical challenge in edge computing. Ma \textit{et al.}~\cite{ma2018efficient} proposed leveraging Docker's layered storage system to minimize file system synchronization overhead, achieving significant reductions in handoff time. The authors in~\cite{10104102} introduce the Mobile Edge Data Handoff (\textbf{MED}) framework for efficient migration of AI inference tasks in edge computing, minimizing accuracy drops and packet loss during handoff. Experimental results demonstrate MED's effectiveness in maintaining inference accuracy for Industry 4.0 applications with proactive data handoff. Similarly, the authors in~\cite{10380573} review Multi-Access Edge Computing (\textbf{MEC}) handover strategies, highlighting algorithms, resource allocation techniques, and challenges in device and application state migration. It identifies gaps in existing research and offers insights to guide future advancements in seamless MEC node transitions.  

\subsection{Microservices Deployment and Resource Optimization}  
Deploying microservices in edge networks with erratic demand patterns has focused on load balancing and resource allocation. Smet \textit{et al.}~\cite{g3} presented a layer placement strategy using iterative heuristics to maximize demand satisfaction while considering response time and storage constraints. Cai~\textit{et al.}~\cite{g8} introduced conTuner, an adaptive tuning framework that enhances container resource configuration, achieving 87\% prediction accuracy in resource contention forecasting. \cite{10639475} addresses performance bottlenecks in edge computing by jointly optimizing microservice deployment and routing using heuristic and reinforcement learning (RSPPO) algorithms. The proposed approach minimizes user delays and resource consumption, achieving load balancing among edge nodes, as verified through extensive experiments. Further, the authors in~\cite{10.1145/3631607} introduce \textbf{Erms}, a resource management system for shared micro-service environments, ensuring SLA compliance with high probability. By profiling latency and optimizing resource scaling, Erms reduces SLA violations by 5× and resource usage by 1.6×, outperforming existing methods in efficiency.

\subsection{Container Performance Evaluation in IoT and Edge Scenarios}  
Bhimani \textit{et al.}~\cite{g10} investigated containerized applications on high-speed NVMe SSDs, presenting a Docker controller that optimized resource utilization and execution time. Morabito \textit{et al.}~\cite{g6} conducted a comprehensive performance evaluation of containerized instances on low-power devices, providing insights for scalable IoT deployments. Alqaisi \textit{et al}.~\cite{g5} compared container technologies, demonstrating Docker's superior performance in memory management and latency-sensitive tasks in ARM-based edge devices. The authors in~\cite{10044213} present a Delay-Aware Container Scheduling (\textbf{DACS}) algorithm for Kubernetes in edge computing, addressing node heterogeneity by considering residual resources and potential delays. Experimental results show that DACS significantly reduces processing and network delays, improving Kubernetes' efficiency in edge environments. Similarly, the authors in~\cite{10743743} introduce a dynamic distributed scheduler for edge computing designed to optimize task execution under diverse IoT constraints such as latency, privacy, and cost. By leveraging predictive profiling and real-time monitoring, the scheduler adapts to dynamic resource and network conditions, demonstrating its effectiveness with an augmented reality application. 

\subsection{Container-Based Solutions for Specific Applications}  
In autonomous driving, Tang et al. \cite{g7} developed a framework for edge offloading with decision-making and scheduling modules, enabling real-time execution with privacy guarantees. Their work underscores the feasibility of container-based solutions for latency-sensitive, resource-intensive applications. The authors in~\cite{10255133} explore telemedicine's reliance on big data pipelines for efficient and secure processing of patient data in distributed settings. It highlights container-based data pipelines within remote patient monitoring systems, addressing challenges like variability and scalability through the cloud/edge/fog continuum, with real-world examples demonstrating practical applications. Further, the authors in~\cite{10688966} propose a container-based platform for emulating Edge computing in smart energy systems, enabling secure and low-latency data processing. A case study on consensus control for multi-micro-grids validates the model's feasibility and highlights its potential for advancing Edge-based energy applications.

\subsection{Motivation} 
Existing research (summarized in Table~\ref{tab:prior_work_pros_cons}) highlights advancements in virtualization, resource management, and task allocation, but significant gaps remain in optimizing interoperability, cost, and resource utilization in distributed IoT ecosystems. This work addresses these challenges by introducing \textbf{AUC-RAC}, a novel framework leveraging Docker swarm networks and auction-based task allocation. By enabling IoT devices to offload computation-intensive tasks to multiple interconnected servers, AUC-RAC optimizes task distribution, enhances resource sharing, and ensures platform independence through containerization. Its profit and cost optimization strategies make it uniquely suited for real-world IoT applications where sustainability and scalability are paramount.

\begin{table}[h!]
\centering
\caption{Summary of Prior Work on IoT, Edge Computing, and Container-Based Solutions.}
\label{tab:prior_work_pros_cons}
\renewcommand{\arraystretch}{1.5}
\resizebox{1.0\textwidth}{!}{
\begin{tabular}{|p{2.5cm}|p{3.8cm}|p{1.5cm}|p{3.0cm}|p{3.0cm}|}
\hline
\textbf{Category} & \textbf{Contribution} & \textbf{Paper} & \textbf{Pros} & \textbf{Cons} \\ \hline

\textbf{Interoperability in IoT Systems} 
& Energy efficiency analysis of Docker; \textbf{IoTvar} middleware simplifies IoT development 
& \cite{g4}, \cite{10.1145/3586010} 
& Reduces developer effort, supports multiple platforms 
& Limited customization, power overhead during operations \\ \hline

\textbf{Virtualization in Real-Time Systems} 
& Impact of Docker on latency; SGP framework; Siemens Industrial Edge performance 
& \cite{g1}, \cite{10406151}, \cite{10144238} 
& Achieves low latency, efficient partitioning, reliable performance 
& Increased computational complexity for SGP, limited scalability \\ \hline

\textbf{Service Handoff in Edge Computing} 
& Efficient service handoff via Docker; \textbf{MED} framework; MEC handover strategies 
& \cite{ma2018efficient}, \cite{10104102}, \cite{10380573} 
& Reduces handoff delays, maintains inference accuracy 
& Packet loss during migrations, high resource demand for MEC algorithms \\ \hline

\textbf{Microservices Deployment and Resource Optimization} 
& Layer placement for demand satisfaction; \textbf{conTuner}; joint deployment-routing optimization; \textbf{Erms} 
& \cite{g3}, \cite{g8}, \cite{10639475}, \cite{10.1145/3631607} 
& Improves SLA compliance, enhances resource efficiency 
& High computational overhead, limited adaptability in erratic networks \\ \hline

\textbf{Container Performance Evaluation in IoT and Edge Scenarios} 
& Docker controller for NVMe SSDs; low-power device evaluation; \textbf{DACS} for Kubernetes; distributed scheduler 
& \cite{g10}, \cite{g6}, \cite{g5}, \cite{10044213}, \cite{10743743} 
& Reduces delays, supports heterogeneous edge environments 
& High initialization time, challenges in predictive scheduling accuracy \\ \hline

\textbf{Container-Based Solutions for Specific Applications} 
& Edge offloading framework; telemedicine pipelines; energy system emulation platform 
& \cite{g7}, \cite{10255133}, \cite{10688966} 
& Supports latency-sensitive applications, ensures secure data processing 
& Complex deployment setups, limited generalizability \\ \hline
\end{tabular}
}
\end{table}

\begin{figure}[h]
\centering
     \includegraphics[scale=.72]{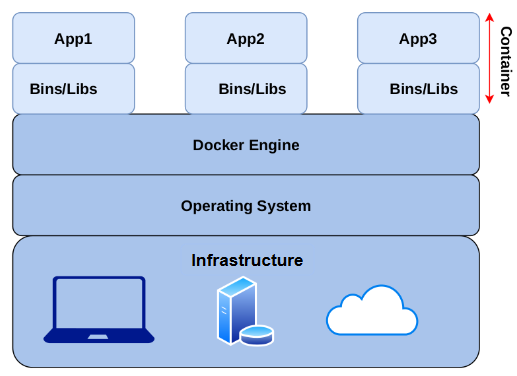}
    \caption{An overview of the layered architecture of the Docker.}
    \label{architecture}

\end{figure}

\section{Preliminaries and problem  statement}\label{applications}
This section presents the preliminaries and problem statements addressed in this work. It starts with the description of the Docker components~\cite{cito2017empirical} followed by Docker Swarm Overview. The section also provides an overview of the layered architecture of the Docker. Docker is a lightweight containerization technique which is used to create, manage, share, and execute applications in Docker containers. It supports the dynamic scale of the application resources, including memory, CPU, \textit{etc}. Docker is an operating system-level virtualization technique. The Docker container was initially based on LXC (Linux container). Fig.~\ref{architecture} illustrates the overview of the layered architecture of Docker, comprising infrastructure, operating system, Docker engine, Bins/Libs, and applications on top.

\begin{figure}[h]
  \centering
  \includegraphics[scale=0.9]{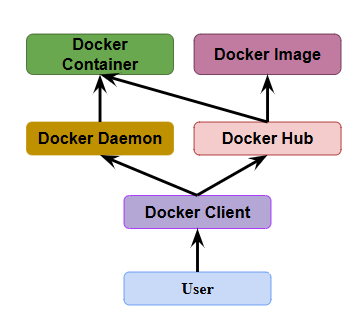}
  \caption{Illustration of the Docker components. }
  \label{taxonomy}
\end{figure}

\subsection{Docker Components}
Docker provides a portable environment for the deployment of software across the container. Various components of Docker, shown in Fig.~\ref{taxonomy}, are discussed as follows.

\begin{itemize}
\item \textbf{Docker client:} The Docker user communicates with the Docker daemon through the Docker client while creating the container, creating images, \textit{etc.} The Docker client uses command and Rest API to communicate with the Docker daemon. The Docker client can interact with one or more Docker daemons.

\item \textbf{Docker-engine/Docker daemon:} The Docker engine runs on the operating system of the host machine. It is responsible for running and managing Docker container services. The Docker user runs some commands, such as Docker run, to start up a container, where the Docker client converts the Docker command into the HTTP API call. Further, it is sent to the Docker daemon, which analyzes the request using the underlying operating system and responds back to the user. It is also responsible for managing images like pulling, pushing, creating an image from a Docker file, \textit{etc.}

\item \textbf{Docker hub:} Docker Hub is an online Docker repository which contains a large number of images. Docker container uses that image with the help of Docker daemon. The Docker container also creates images of the file and sends it to the Docker hub with Docker daemon support.

\item \textbf{Docker Container:} A Docker container is a software module that contains code and all its dependencies files, so the application runs fast and is easy to deploy from one computing environment to another environment. A Docker container image is a lightweight, executable, standalone software module that includes everything required to run an application, like code, system tools, system libraries, and all settings. Any number of containers can be created with the support of a Docker daemon on a single or multiple host machine pool. 

\item \textbf{Docker image}
Docker images are the read-only binary templates used to create a Docker container. In Docker, there are three ways to make the images. 
\begin{enumerate}
    \item \textit{Take the Docker image from the Docker hub} by using the Docker pull command.
    \item \textit{Docker image from the Docker file} In this image creation, the user writes an instruction set in the editor before execution. After execution, files and software groups are required to be placed into a Docker image.
    \item \textit{Create a Docker image from an existing container}.
\end{enumerate}

\end{itemize}

\subsection{Docker Swarm Overview}
Docker Swarm is a container orchestration technique~\cite{g10, moravcik2020overview,ma2018efficient}. It is a service which allows users to create and manage the cluster of the Docker node. It takes multiple Docker daemon running on different host machines, and let’s use them together. Each node of the Docker swarm is a Docker daemon, and all Docker daemons interact using Docker API. Services can be deployed and accessed by the Docker node of the same cluster. Service enables the scale of the Docker application. Docker swarm have two types of nodes: MN and WN. The MN is fully aware of the status of the WN in a cluster. WNs accept tasks given by the MN. Every WN has an agent that reports on the task execution status to the MN. The WN communicates with the manager by using API over HTTP. In a Docker swarm, the service can be deployed and accessed by any node of the same cluster. 

Docker swarm can be used within an IoT environment and requires executing a complex task that is not executed in a single Docker host. The Docker manager receives the task from the IoT device and distributes it to the WNs to execute the task quickly and concurrently. In a Docker swarm, the service can be deployed and accessed by any node of the same cluster. Docker orchestration is created as a task for each service. The task Allocation phase is used to allocate the IP address (of the WN) to the task. The dispatch and scheduler phase is assigned and instructs nodes to run a task. 

\subsection{Problem statement}
In this work, we target to solve the following problem: \textit{How to leverage both auction-based bidding and containerization techniques to enhance cost efficiency and execution effectiveness in IoT environments?} The proposed approach utilizes a distributed architecture with a master node responsible for receiving tasks and worker nodes to execute the task. The distinctive feature of our approach is its dual-level cost optimization approach, which is discussed in the next section.

\section{\textbf{AUC-RAC}: \textbf{AUC}tion-based mechanism for task allocation and \textbf{R}esource \textbf{A}ware \textbf{C}ontainerization} \label{sec5}
\begin{figure}[h]
\centering
\includegraphics[scale=1.1]{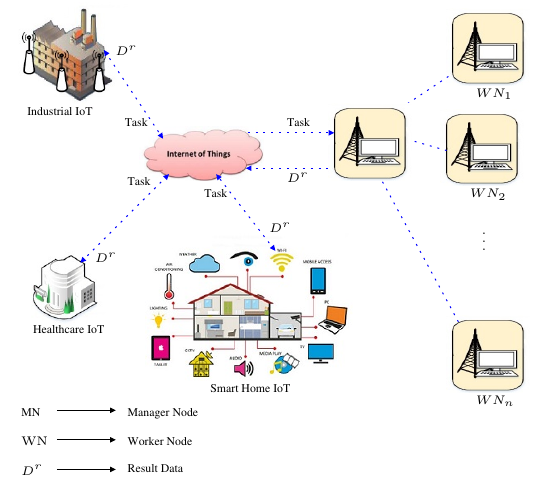}
\caption{An overview of AUC-RAC approach.}
\label{overview}
\end{figure}

This section proposes an auction-based mechanism for task allocation and resource-aware containerization, acronyms as AUC-RAC. The objective of the proposed mechanism is to execute the IoT-driven task within the given deadline while ensuring cost optimization. AUC-RAC involves distributed architecture comprising \textbf{Master Node (MN)} and \textbf{Worker Nodes (WNs)}, where the task-receiving node is termed as the MN and all other nodes involved in the task execution are called WNs. Apart from the existing work, we assume all the WNs use containerization techniques to curtail the time and cost of execution. Through this containerization, we propose \textit{one of its kind approaches}, where we ensure cost optimization at \textbf{task distribution level} and at the \textbf{task execution level}. To ensure cost optimization at the task distribution level, we use \textbf{auction-based bidding techniques}, which is part of the Bayesian game. Further, we use \textbf{resource aware containerization} to ensure cost optimization at the execution level. Fig.~\ref{overview} illustrates the proposed AUC-RAC approach.

AUC-RAC approach starts with cost optimization at the task distribution level using a game theory-based bidding technique, where the auctioning agent (\textit{i.e.,} MN) allocates the portion of IoT task among competitive WNs based on their submitted bid value, detailed discussed in Section~\ref{tdl}. Next, each WN performs cost optimization using resource-aware containerization to the sub-fraction of the allocated task portion to execute the task using multiple containers, as discussed in Section~\ref{rac}. Algorithm~\ref{algorithm1} summarizes the different steps of the AUC-RAC approach.  

\subsection{Cost optimization at task distribution level}\label{tdl}
This subsection describes the cost optimization at the task distribution level, where we opt for the bidding-based auction technique of game theory. The section explores the participation behaviour of individuals in auction processes and examines the dynamics of goods in auction markets using a bidding-based auction model in economics. To this end, we introduce several fundamental concepts, including definitions and theorems, that underpin the mechanics of auction techniques. The auctioning agent (or MN) allocates IoT tasks among competitive WNs based on their submitted bids. Moreover, the bid value of each WN depends on the personalized assessment, \textit{e.g.,} \textit{calculation of task execution cost based on the task details received from the MN}. Each WN performs its assessment as follows:

\subsubsection{WN personalized assessment}\label{sec5.11}
Each work node performs some personalized assessment or internal valuation for a task before submitting the bid to the MN. The assessment calculation is private and follows the WN resources and policies. This calculation helps the WN to submit its optimal bidding price, which is further utilized by the MN to allocate the task portion for execution. Now, we define the different parameters considered by each WN to estimate its personalized assessment. Let us consider a set $W=\{w_1,\cdots, w_n\}$ of $n$ WNs in this work; they act as the local and distributed server for task execution. Among $n$ WNs, only a few satisfy the selection criteria for task execution based on the bid value. Further, we assume the computation capability of a WN $i$ as $e_i$ cycles per second and the available memory on WNs is $m_i$, where $i\in \{1,\cdots, n\}$. 

Let set of task $T = \{t_1,\cdots, t_m\}$, where $j=\{1,\cdots,m\}$, is to be executed on the workers nodes. The end user wants to execute task $t_j$ on a given time instance having a data size of $d_j$. The MN transfers the task $t_j$ to a WN $w_i$ and gets the result of the size $d_j^r$. Let the execution cost of $t_j$ on $w_{i}$ be $c_{ij}$ that requires $e_{ij}$ cycles of CPU and $m_{ij}$ MB for its execution on $w_i$, consuming power $p_{ij}$ watts. 

Further, assuming the required resource $R_j$ for executing $t_j$, which encompasses CPU cycles, memory, and power, is a function defined as follows:
\begin{align}\nonumber
\mathcal{F}^{res}_j(e_j, m_j, p_j) & =\delta(R_j)\\\label{resource}
&= \delta(\lambda_1 e_j+ \alpha_1\lambda_2 m_j + \alpha_2\lambda_3 p_j),
\end{align}
where $\delta$ is the mapping constant that maps function $\mathcal{F}^{res}_j(\cdot)$ with $R_j$. $\lambda_1, \lambda_2,$ and $\lambda_3$ are the variables that manage the contribution of each parameter. For simplicity, we consider $\lambda_1=\lambda_2=\lambda_3=1/3$. $\alpha_1$ and $\alpha_2$ are the unit matching factors that manage the mismatch between different units of memory and power in contrast with unit-less quantity CPU cycles. 

From Eq.~\ref{resource}, we estimate the internal valuation computed by each $w_i$, where $i\in\{1,\cdots,n\}$, in the form of execution cost $c_{ij}$. The cost $c_{ij}$ is estimated as:  
\begin{align}\nonumber
c_{ij} \propto &\mathcal{F}^{res}_{ij}(e_{ij}, m_{ij}, p_{ij}), \\\label{cost}
\implies  c_{ij} = &c_i \mathcal{F}^{res}_{ij}(e_{ij}, m_{ij}, p_{ij}),
\end{align}
where $c_i$ is the proportionality cost or cost required per unit on $w_i$. The estimated cost in Eq.~\ref{cost} helps the WN to determine the auction bid.

WNs participate in a sealed bid process by submitting bids determined through the internal valuation of task cost. Sealed bid auctions, both the initial and subsequent ones, are frequently employed in auction-based mechanisms. These bidding techniques are integral to the \textit{Bayesian game}, characterized by incomplete information~\cite{wu2021agent}. In this context, numerous IoT devices are connected to a central server (MN), and MN makes task offloading decisions among multiple local servers (WNs). The coordination between the MN and WNs is facilitated by utilising Docker swarm concepts.

To compensate for task execution, the MN remunerates the WNs based on the agreed-upon prices. Notably, each WN can concurrently execute multiple tasks with diverse platform dependencies. This is made possible through the implementation of Docker containerization techniques, allowing for the seamless execution of IoT tasks. Further, Docker containers facilitate updates or removal of functionalities based on user requirements, enhancing adaptability and efficiency.

\subsubsection{WN profit maximization}
The objective of profit maximization for each WN is achieved through the computation of a cost function in Eq.~\ref{cost}. The valuation, estimated through the cost, serves as a critical determinant in the bidding process. WNs strategically assess the inherent value and complexity of the task, aiming to enhance their profitability by submitting competitive bids to MN. The cost function, thus, encapsulates the important considerations and valuation metrics employed by each worker. Using Eqs.~\ref{resource} and \ref{cost}, final cost is obtained as:
\begin{align} 
c_{ij} &= c_i \delta(\lambda_1 e_{ij}+ \alpha_1\lambda_2 m_{ij} + \alpha_2\lambda_3 p_{ij}),\\\label{cf}
& = c_i \delta\Big\{\lambda_1 \Big(\frac{e_{j}}{e_i}\Big)+ \alpha_1\lambda_2 \Big(\frac{m_{j}}{m_i}\Big) + \alpha_2\lambda_3 \Big(\frac{p_{j}}{p_i}\Big)\Big\}.
\end{align}

The goal of a WN is to minimise the cost $c_{ij}$ (in Eq.~\ref{cf}) to maximize their profit. Thus, the minimization problem is defined using the following expression:
\begin{subequations}\label{main}
\begin{align}
  \operatorname{min}_{e_j,m_j,p_j} & \quad c_{ij}\\
 = c_i & \delta\Big\{\lambda_1 \Big(\frac{e_{j}}{e_i}\Big)+ \alpha_1\lambda_2 \Big(\frac{m_{j}}{m_i}\Big) + \alpha_2\lambda_3 \Big(\frac{p_{j}}{p_i}\Big)\Big\},
\end{align} 
\qquad Subject to:
\begin{align}
\lambda_1+\lambda_2+\lambda_3=1,\\
\lambda_1, \lambda_2, \lambda_3 < 1,\\
\frac{e_{j}}{e_i}, \frac{m_{j}}{m_i}, \frac{p_{j}}{p_i}< 1,\\
\Omega^{req}_j \le \Omega^{max}_j,
\end{align}
\end{subequations}
where $\Omega^{req}_{ij}$ and $\Omega^{max}_j$ are the time required to execute the task $t_j$ on $w_j$ and the deadline (or maximum allowable time) for executing the task $t_j$, respectively. We can define $\Omega^{req}_{ij}$ as the function of processing power or required CPU cycle using a time constant $\phi_i$, as:
\begin{equation}\label{time}
\Omega^{req}_{ij} = \phi_i \frac{e_{j}}{e_i}.
\end{equation}

Next, we apply Lagrangian to obtain the solution of the optimization problem in Eq.~\ref{main}:\\

\noindent \textbf{Lagrangian:}
\begin{flalign}\nonumber
L(e_j, m_j,& p_j,\lambda_1, \lambda_2, \lambda_3, \lambda_4) = (\lambda_1 e_i + \lambda_2 \alpha_1 m_i + \lambda_3 \alpha_2 p_i) \\ \nonumber
&\cdot \left\{\lambda_1 \left(\frac{e_{j}}{ e_i}\right) + \lambda_2\left(\frac{m_{j}}{m_i}\right) + \lambda_3\left(\frac{p_{j}}{ p_i}\right)\right\} \\ \nonumber
&+ \lambda_1(\lambda_1 + \lambda_2 + \lambda_3 - 1) + \lambda_2(\lambda_1 + \lambda_2 + \lambda_3 - 1) \\ \nonumber
&+ \lambda_3(\lambda_1 + \lambda_2 + \lambda_3 - 1) + \lambda_1^2\left(1 - \frac{ e_{j}}{ e_i}\right)\\ \nonumber
&+ \lambda_2^2\left(1 - \frac{m_{j}}{m_i}\right) + \lambda_3^2\left(1 - \frac{p_{j}}{p_i}\right) \\ \label{lag}
&+ \lambda_4\left(\phi_i\frac{e_j}{e_i} - \Omega^{max}_j\right).
\end{flalign}

\noindent \textbf{Partial Derivatives:}
\begin{align}\nonumber
\frac{\partial L}{\partial e_j} = 0, \frac{\partial L}{\partial m_j} = 0, \frac{\partial L}{\partial p_j} = 0\\
\frac{\partial L}{\partial \lambda_1} = 0, \frac{\partial L}{\partial \lambda_2} = 0, \frac{\partial L}{\partial \lambda_3} = 0
\frac{\partial L}{\partial \lambda_4} = 0.
\end{align}

Solving Eq.~\ref{lag} provides the values for \(e_j\), \(m_j\), \(p_j\), \(\lambda_1\), \(\lambda_2\), \(\lambda_3\), and \(\lambda_4\) that minimize the linear cost function while satisfying the constraints. However, finding an analytical solution for this system of equations might be complex due to the nonlinear nature of the problem. As a result, numerical methods such as iterative optimization algorithms might be more suitable for obtaining the solution. To simplify the solution, we use $\lambda_1+\lambda_2+\lambda_3=1$ and $\lambda_1 = \lambda_2=\lambda_2=1/3$ in Eq.~\ref{lag}, as it align with our constraints. Thus, we obtain :
\begin{flalign}\nonumber
L(e_j, m_j,& p_j,\lambda_1, \lambda_2, \lambda_3, \lambda_4) = \frac{1}{3}(e_i + \alpha_1 m_i + \alpha_2 p_i) \\ \nonumber
&\cdot \left\{\left(\frac{e_{j}}{3e_i}\right) + \left(\frac{m_{j}}{3m_i}\right) + \left(\frac{p_{j}}{3p_i}\right)\right\} \\ \nonumber
& \frac{1}{9}\left[\left(1 - \frac{ e_{j}}{ e_i}\right) + \left(1 - \frac{m_{j}}{m_i}\right) + \left(1 - \frac{p_{j}}{p_i}\right) \right]\\ \label{lag2}
&+ \lambda_4\left(\phi_i\frac{e_j}{e_i} - \Omega^{max}_j\right).
\end{flalign}

\begin{algorithm}[t]
\caption{\textbf{WN cost optimization algorithm.}}
\label{algo2}
\KwIn{$e_i, m_i, p_i, \alpha_1, \alpha_2, \phi_i, \Omega^{max}_j, e_j, m_j, p_j$;}
Using Lagrangian in Eq.~\ref{lag2} and obtain  its \textbf{partial derivatives}\;
\medskip
/*Set up the system of equations*/\\
\medskip
\For{partial\_derivative $\in$ \textbf{partial derivatives}}
    {
     system\_of\_equations = equation(partial\_derivative, 0)
    }
\medskip
/*Solve the system of equations for critical points*/
\medskip
critical\_points = \textbf{solve}(system\_of\_equations, $(e_j, m_j, p_j)$)
\textbf{return} $e_j, m_j, p_j$\;
\medskip
Function \textbf{solve}(system\_of\_equations, $(e_j, m_j, p_j)$):\;
\hspace{0.5cm} Set initial values for the variables\;
\hspace{0.5cm} learning rate $\psi$\;
\While{not converge}
{
Evaluate partial derivatives at the current values\;
Update each variable = variable - $\psi\times$ partial\_derivative\;
Adjust the $\psi$ based on the convergence behaviour \;
\If{convergence criteria}{
\textbf{break}\;
}
\textbf{return} the final values of $e_j, m_j, p_j$\;
}   
\end{algorithm}

To address the optimization problem presented in Eq.~\ref{lag2}, Algorithm~\ref{algo2} outlines a fundamental iterative optimization approach. The practical implementation necessitates the selection of a suitable optimization method, specification of convergence criteria, and handling of mathematical operations using the chosen programming language. The algorithm encompasses various steps, including variable definition, computation of partial derivatives, establishment of a system of equations, solution of the system, and, ultimately, the return of critical points. Example~\ref{ex1} depicts an example scenario of the proposed algorithm for solving the optimization problem in Eq.~\ref{lag2}. 

\begin{example}\label{ex1}
Let us consider arbitrary values for the parameters \(e_i\), \(m_i\), \(p_i\), \(\alpha_1\), \(\alpha_2\), \(\phi_i\), and \(\Omega^{max}_j\). We will solve for the critical points with respect to \(e_j\), \(m_j\), \(p_j\), \(\lambda_1\), \(\lambda_2\), \(\lambda_3\), and \(\lambda_4\) in Eq.~\ref{lag2}. Assuming arbitrary values: $e_i = 2, \quad m_i = 3, \quad p_i = 4, \quad \alpha_1 = 0.5, \quad \alpha_2 = 0.8, \quad \phi_i = 10, \quad \Omega^{max}_j = 15.$
Now, computing the partial derivatives and set them to zero: $\frac{\partial L}{\partial e_j} = 0, \frac{\partial L}{\partial m_j} = 0, \frac{\partial L}{\partial p_j} = 0, \frac{\partial L}{\partial \lambda_1} = 0, \frac{\partial L}{\partial \lambda_2} = 0, \frac{\partial L}{\partial \lambda_3} = 0, \frac{\partial L}{\partial \lambda_4} = 0$. Upon simplifying and solving the system of equations, the critical points are $e_j = \frac{4}{3}, \quad m_j = \frac{2}{3}, \quad p_j = \frac{8}{3}, \quad \lambda_1 = -\frac{7}{18}, \quad \lambda_2 = \frac{1}{6}, \quad \lambda_3 = -\frac{5}{18}, \quad \lambda_4 = 0.$ Notably, these values are specific to the assumed parameters \(e_i\), \(m_i\), \(p_i\), \(\alpha_1\), \(\alpha_2\), \(\phi_i\), and \(\Omega^{max}_j\). If we have different values for these parameters, the critical points change accordingly.
\end{example}

\subsection{Cost optimization at task execution level}\label{rac}
In this subsection, we discuss the mechanism of cost optimization at the task execution level after the discussion of cost optimization at the task distribution level in Section~\ref{tdl}. After receiving tasks from IoT devices, the MN dispatches task details to all WNs. Numerous WNs actively participate in the bidding process and submit their bids to the MN. All eligible nodes take part in the auction, with eligibility determined by the resource availability of the WNs. Bidding-based auctions in economics provide insights into how individuals engage in the auction process and assess goods within auction markets. 
The allocation of goods, represented by IoT tasks in this scenario, occurs among competitive WNs based on their bids in the auction. The bid value is contingent upon the private calculation of task execution costs by WNs, which is influenced by the specific details of the task. 

MN charges $d_j \times v$ price to the IoT client for the task execution, where $v$ represents the unit cost charged by the MN to the client, and $d_j$ signifies the data transfer from the IoT client to the MN for the execution of task $t_j$. We utilise private valuations cost, $c_{ij}$ (in Eq.~\ref{cf}), for each WN $i$ corresponding to $t_j$. The computation of private valuations for task nodes remains confidential and is exclusive to the respective agent.

Apart from task allocation, an auction also establishes the payments for the agent that successfully executes the task. The function $\mathcal{F}^{res}_j(e_j, m_j, p_j) =\delta(R_j)$ (in Eq.~\ref{resource}) decides the bidding value of the WN $i$. We denote $b_{i,j}$ as the bidding value submitted by WN $i$ for task $j$ and $b_{-i,j}$ as all bids except $b_{i,j}$ for task $j$. Let task $t_j$ possess a task execution deadline of $r_j$ and $q_{ij}$ be the time taken by WM $i$ to execute $t_j$. Thus, WM $i$ participate in bidding if it satisfies the following condition:

\begin{equation}\label{zeta}
\zeta_{ij}=\left\{\begin{matrix}
1 & (r_j-q_{ij})>0, \\ 
0 & \text{otherwise}.
\end{matrix}\right.
\end{equation}

Further, the chance of getting the task $t_j$ on a given bid value $b_{i,j}$ is a random event. Thus, we derive the mathematical expression for the probability of winning bids, which typically involves the distribution of bids and the criteria for winning. For simplicity, we assume a simple scenario with sealed-bid auctions where the highest WN or WN wins.  

Let $F(b_{i,j})$ be the cumulative distribution function (CDF) of the bids. The probability of winning the bid can be expressed mathematically as the probability that a WN submits the highest bid. If $b_{ij}$ is the bid submitted by WN $i$ and $b_{kj}$ is the bid submitted by another WN $k$ for task $t_j$, then the probability of winning WM $i$ can be expressed as:
\begin{equation}
P_{win}^i = P(b_{i,j} > b_{-i,j}), i\in\{1,2,\cdots,n\}.
\end{equation}

This can be further expanded using the joint probability:
\begin{equation}
P_{win}^i = \int_{-\infty}^{\infty} P(b_{i,j} > b \ \text{and} \ b_{-i,j} \leq b \ \forall \ k \neq i) \cdot f(b)\, db,
\end{equation}
where \(f(b)\) is the probability density function (PDF) of the bids. The integral takes the entire range of possible bid values.

Since there are $N$ WNs, let us derive the mathematical expression for the probability of winning bids in an auction with $n$ players. Assuming independent and identically distributed bids, the probability of winning for a specific WN (considering WN $i$) is given by:
\begin{equation}
P_{Win}^i = \left[ \int_{-\infty}^{\infty} P(X_i > b) \cdot f(b) \,db \right]^{n-1}. 
\end{equation}

We assume that the bids are independent and identically distributed, and the integral is taken over the entire range of possible bid values. This expression captures the joint probability of all other players submitting bids less than or equal to the specific WN $i$'s bid, and it is raised to the power of $n-1$ to account for all other players. For example, if bidding price $b_{1,j}> b_{2,j}$ for task  $t_j$, then WM $1$ win auction for $t_j$. 

Further, we determine the mathematical expression for profit of the WN. Let us consider a scenario where a WN $i$ wants to determine its bid value and the associated winning probability in the auction. In a sealed-bid auction, the highest WN wins. The goal is to find the bid value $b_{i,j}$ for executing task $t_j$, which maximizes the WN's expected utility.

WN's utility ($U_{ij}$) can be modelled as the difference between the perceived value, let's say $V_j$, and the bid submitted $b_{i,j}$ if the WN $i$ wins for executing task $t_j$. If WN loses, the utility is zero.
\begin{equation}
U_{ij}(b_{i,j}) = \begin{cases} V_j - b_{i,j}, & \text{if } b_{i,j} > b_{-i,j}\ \forall \ i \in \{1,2,\cdots,n\} \\ 0, & \text{otherwise} \end{cases}.
\end{equation}
The expected utility involves winning probability, utility, and deadline constraint $\zeta_{ij}$ (Eq.~\ref{zeta})  can be expressed as:
\begin{equation}\label{expect}
\text{E}[U_{ij}(b_{i,j})] = P_{Win}^i \cdot (V_j - b_{i,j}) \cdot \zeta_{ij}.
\end{equation}

\noindent $\bullet$ \textbf{Utility maximization problem of WN:}
We are interested in finding the bid $b_{i,j}$ that maximizes this expected utility. Therefore, we seek to maximize the expression in Eq.~\ref{expect} of the expected utility by finding the optimal bid $b_{i,j}$. WN's optimization problem is given as:
\begin{subequations}
\begin{align}
\max_{b_{i,j}}  & \left\{ P_{Win}^i) \cdot (V_j - b_{i,j})   \cdot \zeta_{ij} \right\},\\ \nonumber
\text{subject to:} & \\
& V_j - b_{i,j} \ge 0,\\
& r_j-q_{ij}\ge 0,\\
& 1\le j \le m \text{ and } 1\le i \le n.
\end{align}
\label{opti1}
\end{subequations}
\noindent $\bullet$ \textbf{Solution:} This involves taking the derivative concerning $b_{i,j}$, setting it to zero, and solving for the optimal bid. The exact form of the solution depends on the specifics of the bid distribution and the WN's perceived value $V_j$ for task $t_j$. The solution may involve elements such as the bid distribution's PDF, CDF, and the number of WNs.

To solve the maximization problem, we differentiate the expected utility function concerning the bid $b_{i,j}$, set the derivative to zero, and solve for the optimal bid. Let us denote the optimal bid as $b_{i,j}^*$.
\begin{equation}
\frac{d}{db_{i,j}}\left\{ P_{Win}^i \cdot (V_j - b_{i,j}) \right\} \cdot \zeta_{ij}= 0.
\end{equation}

Now, we focus on finding the optimal bid for WN $i$ under the assumption of independent and identically distributed bids. The winning probability \(P(\text{Win}_i)\) is given by:
\begin{equation}
\frac{d}{db_{i,j}}\left\{ \left[ \int_{-\infty}^{\infty} P(b_{i,j} > b) \cdot f(b) \,db \right]^{n-1} \cdot (V_j - b_{i,j}) \cdot  \zeta_{ij}\right\} = 0.
\end{equation}

To simplify the expression, let us denote $P_i = \int_{-\infty}^{\infty} P(b_{i,j} > b) \cdot f(b) \,db$. Then:
\begin{equation}
\frac{d}{db_{i,j}}\left\{ P_i^{n-1} \cdot (V_i - b_{i,j}) \cdot \zeta_{ij} \right\} = 0.
\end{equation}

Now, differentiating the above equation:
\begin{equation}
(n-1) \cdot P_i^{n-2} \cdot \frac{dP_i}{db_{i,j}} \cdot (V_i - b_{i,j}) \cdot \zeta_{ij} - P_i^{n-1} \cdot \zeta_{ij}= 0.
\end{equation}

Simplifying and solving for \(\frac{dP_i}{db_{i,j}}\):
\begin{equation}
\frac{dP_i}{db_{i,j}} = \frac{P_i}{(n-1) \cdot (b_{i,j} - V_j)}.
\end{equation}

Now, substitute this back into the original equation:
\begin{equation}
\frac{P_i}{(n-1) \cdot (b_{i,j} - V_j)} \cdot (V_j - b_{i,j}) - P_i = 0.
\end{equation}

Simplifying the above equation:
\begin{equation}
\frac{V_j - b_{i,j}}{(n-1) \cdot (b_{i,j} - V_j)} = 1.
\end{equation}

\begin{equation}
\implies (V_j - b_{i,j}) = (n-1) \cdot (b_{i,j} - V_j).
\end{equation}

Expanding and solving for $b_{i,j}$:
\begin{equation}
V_j - b_{i,j} = (n-1) \cdot b_{i,j} - (n-1) \cdot V_j.
\end{equation}
\begin{equation}
n \cdot b_{i,j} = n \cdot V_j.
\end{equation}
\begin{equation}
b_{i,j} = V_j.
\end{equation}

Thus, under the assumption of independent and identically distributed bids, the optimal bid for WN $i$ to maximize their expected utility is equal to their received value $V_j$. It indicates that the WNs now commence competition among themselves to decrease their bid values, reducing them as close to $V_j$. The WN whose bidding reaches \(V_j\) first will be assigned the tasks \(t_j\); different steps are discussed in Algorithm~2.

\begin{algorithm}[t]
\caption{\textbf{Task allocation to Worker Nodes.}}
\label{algo3}
\KwIn{List of received bids from WNs $\mathbf{V}=\{V_1,V_2,\cdots,V_n\}$ and list of tasks to be assigned $T=\{t_1,t_2,\cdots,t_m\}$;}
Sort the WNs in ascending order of values: $\mathbf{V} \gets \text{sort}(\mathbf{V})$\;
Initialize bids for each WN: $\mathbf{bids} \gets [0] * \text{len}(\mathbf{V})$\;
\For{each task $t_j$ in the list of tasks $\mathbf{T}$}
    {
    \medskip
    /*Find the index of the first WN whose bid reaches or exceeds the task value*/\\
    \medskip
         $WN\_index \gets$ index of first WN with bid $\geq$ task value\;
         \If{$WN\_index$ is \textbf{None}}
            {$WN\_index \gets \text{len}(\mathbf{V}) - 1$\;
            \medskip
            \nonl /*Assign to the WN with the highest bid*/\\
    }
    }
Assign the task $t_j$ to the WN with index $WN\_index$\;
\medskip
/*Update the bid for the assigned WN:*/ \\
\medskip
$bids[WN] \gets \max(bids[WN], \text{task\_value})$\;
   
   \textbf{Output:} List of optimal bid values for each WN: $\mathbf{bids}$
\end{algorithm}

\subsection{Resource allocation to the container}
Resource allocation is a critical aspect of the containerization technique, as it directly impacts the profitability of Worker Nodes (WNs) during task execution. Each WN hosts multiple containers to execute tasks, and selecting and creating containers in the Docker environment is vital. In this section, we determine the appropriate Docker container and its size based on resource requirements. Let $td_j^{max}$ represent the maximum time delay for executing task $j$. Based on the function $f$, we select a container with the minimum available resources. The system offers multiple containers, and selection is made using a best-fit algorithm.

The container resources must accommodate the required library installations, defined as $(c_i + m_i)$, where $c_i$ is the memory resource needed for library installation. A local server will create a container if no existing containers have the necessary resources or if no idle containers are available. Random resource allocation leads to inefficient server usage and reduces task execution concurrency. The resource allocation steps are summarized in Algorithm~3.
       
\begin{algorithm}[h]
\SetAlgoLined
\textbf {Input}: Task $j$ available in the WN $i$ after winning the bid\;
\KwResult{Chosen container and allocate task $j$.}
\While{true}{
\textbf{re: }\eIf{container available}{
   Search the free container\;
   Arrange the obtained free containers in increasing order of their computation power and memory\;
   a[] $\leftarrow$ free container id \;
    n $\leftarrow$ a.size() \;
    sort(a) and initialize $i=0$ \;
      
        \While{$n > i$ }{
          
           \eIf{$(m_c > m_j)$ and $(\frac{C_j}{C_c} < td_j^{max})$}{
            
           container id $\leftarrow$  a[i] \;
           allocate task j \;
            i = i+1\;
            break \;
   }{
   i = i+1\;
  }
        } 
   }{
   
   \eIf{$(m_i > m_j) and (\frac{C_j}{f_i} < td_j^{max})$}{
   Create container \;
     $ m_C \leftarrow m_j$\;
    $allocate \rightarrow C_c \Rightarrow$  $(\frac{C_j}{C_c} < td_j^{max})$ \;
   }{
   Goto \textbf{re}\;
  }
  
  }
 }
 \caption{\textbf{Resource allocation to the container.}}
 \label{algorithm1}
\end{algorithm}

\begin{figure}[h]
\centering
    \includegraphics[height=4cm,width=8.5cm]{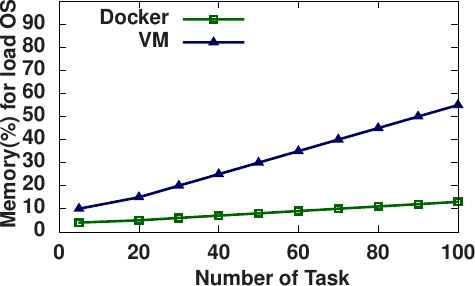}
    \caption{Impact of task count on the memory for loading operating system (OS). }
    \label{fig1}
 \end{figure}

\section{Experimental results}\label{experiment}
The proposed auction-based mechanism for efficient computation task offloading among multiple local servers in the context of IoT devices was comprehensively evaluated and yielded promising outcomes. The experiments were conducted using a simulated environment that mimics real-world IoT scenarios. The evaluation focused on assessing the fairness of task allocation, the tolerable delay experienced by IoT devices, and the overall profit maximization of the management system. 

\begin{figure}
\centering
    \includegraphics[height=4cm,width=8.5cm]{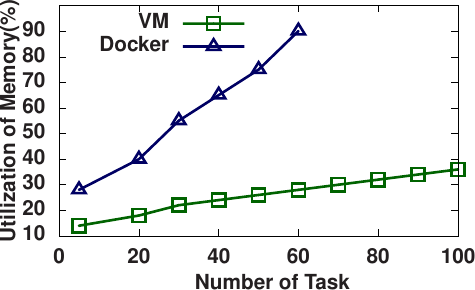}
        \caption{Impact of task count on the memory utilization.}
    \label{fig2}
 \end{figure}

\begin{figure}
\centering
    \includegraphics[height=4cm,width=8.5cm]{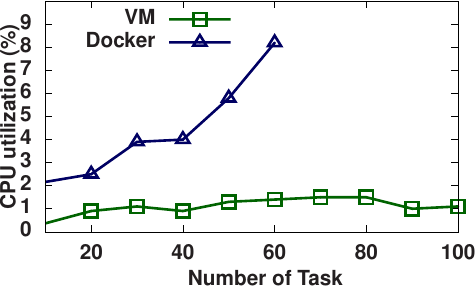}
    \caption{Impact of task count on the CPU utilization.}
    \label{fig3}
\end{figure}

The study compares resource allocation techniques for virtual machines, focusing on memory utilization. It finds that Docker outperforms traditional virtual machines, especially as the number of parallel executions increases, due to its ability to leverage the host machine's operating system. This efficiency is further enhanced in Docker Swarm, which improves resource management in clustered environments. In tests involving over a hundred concurrent IoT tasks executed in a Docker container using a master-slave system, Docker Swarm containerization surpasses other virtualization techniques. These findings, depicted in Figs.~\ref{fig1}, \ref{fig2}, and \ref{fig3}, highlight Docker's superior scalability and efficiency, making it an optimal choice for applications requiring effective resource management.

The auction-based bidding process effectively ensures fairness among participating local servers, as demonstrated in the distribution of computation tasks (Table~\ref{tab1}). Each server had an equitable opportunity to engage in task execution, leading to a consistently balanced workload across the network. This approach prevents any single server from monopolizing resources, promoting a just and unbiased allocation system. The efficacy of this mechanism is highlighted in Table~\ref{tab1}, which illustrates the even distribution of tasks among servers, supporting the fairness and efficiency of the proposed method. The auction-based system optimises overall network performance and reliability by dynamically assigning tasks based on current server capabilities.

\begin{table}[h]
\centering
\caption{Illustration of task distribution in the proposed approach. LIT, MIT, and HIT are Low, Medium, and High-intensity tasks, respectively.}
\begin{tabular}{|l|c|c|c|}
\hline
\textbf{Strategy }                & \textbf{\% LIT} & \textbf{\% MIT} & \textbf{\% HIT} \\ \hline
Static Offloading        & 30                        & 40                           & 30                         \\ \hline
Heuristic Offloading     & 20                        & 50                           & 30                         \\ \hline
Auction-Based Offloading & 40                        & 30                           & 30                         \\ \hline
VM Offloading            & 25                        & 35                           & 40                         \\ \hline
\end{tabular}
\label{tab1}
\end{table}

\begin{figure}
\centering
 \includegraphics[height=4cm,width=8.5cm]{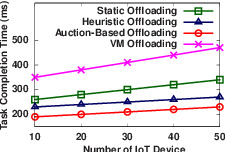}
  \caption{Task completion time \textit{vs.} number of IoT devices.}
  \label{master21}
\end{figure}

In IoT applications, one of the crucial factors is the delay encountered by devices when offloading and executing tasks. Managing these delays is vital to ensure smooth and efficient operation. The experimental results showed that the proposed system adeptly kept delays within acceptable limits, as illustrated in Fig.~\ref{master21}. By employing an auction-based approach, the system intelligently prioritized the allocation of tasks, leading to minimal delays for IoT devices. This approach ensured that tasks were assigned based on current network conditions and device capabilities, optimizing overall execution time. Consequently, the system successfully balanced efficient resource utilization with the need for prompt task completion, demonstrating its capability to enhance IoT performance by reducing unnecessary latency and improving response times.

The analysis presented in Table~\ref{tab3} delves into the management system's profit, which serves as a key indicator of the auction-based mechanism's success. This model consistently achieved a significant increase in profit compared to traditional offloading strategies. By strategically allocating computation tasks based on available resources and incorporating a competitive bidding process, the proposed approach significantly enhanced the system's overall profitability. This outcome highlights the economic viability and sustainability of the auction-based mechanism in practical applications.

To assess the system's scalability and performance, the evaluation was extended to scenarios involving different numbers of IoT devices and local servers. The results showed that the proposed mechanism maintained its efficiency and effectiveness as the network scaled up. This robust scalability indicates the system's capacity to support IoT ecosystems of various sizes and complexities without sacrificing its core objectives of fairness, minimizing delay, and maximizing profit.

\begin{figure}
\centering
    \includegraphics[height=6cm,width=11cm]{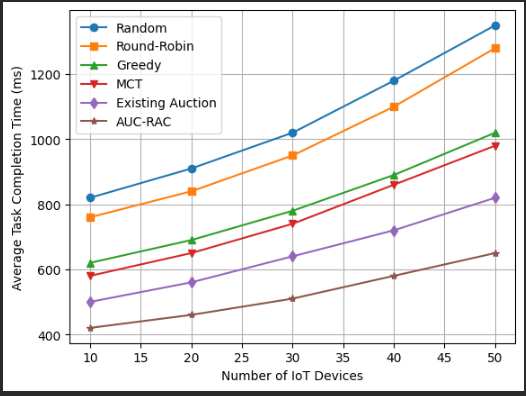}
    \caption{Impact of task count on the CPU utilization.}
    \label{rev1}
\end{figure}

As the number of IoT devices in the system increases, In figure \ref{rev1}, shows the variation of the average job completion time. Our proposed AUC-RAC framework is compared with the baseline task allocation strategies like : Random, Round-Robin, Greedy, Minimum Completion Time (MCT), and an existing auction-based technique. The number of IoT devices varies between 10 and 50.
We observed from the figure that the Random and round robin techniques are the highest task completion time across all scales. This is basically due to their lack of resource awareness an inability to adapt to heterogeneous server capabilities, to result in ineffective task placement and increased execution delay time.
The Greedy and MCT methods achieve shorter completion times compared to Random and Round- Robin by taking resource availability during task assignment. However, these approaches fail to take container-level constraints into consideration, which lead results in suboptimal performance.
The existing auction-based approach further degrades task execution time by employing competitive bidding among servers, allowing better resource selection. Nevertheless, its performance declined at higher device counts due to the absence of container-aware resource management.
In contrast, our proposed AUC-RAC approach achieves the lowest average task completion time for all system sizes. This enhancement is attributed to the combined integration of auction-based task allocation, resource awareness, and container-level constraints, which allow efficient and scalable task completion in a distributed IoT system.
Overall, our results show that AUC-RAC significantly improves the performance compared to existing methods, particularly under high system load. This validates its effectiveness and scalability for computationally intensive IoT applications.

The auction-based mechanism developed in this research offers an innovative solution for efficiently offloading computation tasks among multiple local servers in IoT environments. The results demonstrate its ability to ensure fairness among servers, reduce delays for IoT devices, and enhance the profitability of the management system. Leveraging Docker Swarm and containerization further improves the system's versatility and adaptability. This study underscores the considerable potential of distributed computing paradigms to boost the performance and efficiency of IoT devices through collaborative resource sharing.

We clearly demonstrated the novelty and design advantage of our proposed AUC-RAC strategy, and we compared it with various existing baseline task-allocation techniques commonly used in distributed frameworks or IoT edge computing frameworks. These existing methods are selected to reflect various level of system awareness, including resource awareness , auction-based allocation, container awareness, and distributed task execution. In Table~\ref{rev2} we present a comparative analysis of some existing methods based on key attributes.

\begin{table}[t]
\centering
\caption{Comparative analysis of existing Task Allocation methods}
\label{tab:baseline_comparison}
\begin{tabular}{lcccc}
\hline
\textbf{Method} & \textbf{Resource-Aware} & \textbf{Auction-Based} & \textbf{Container-Aware} & \textbf{Distributed} \\
\hline
Random & $\times$ & $\times$ & $\times$ & $\checkmark$ \\
Round-Robin & $\times$ & $\times$ & $\times$ & $\checkmark$ \\
Greedy & $\checkmark$ & $\times$ & $\times$ & $\checkmark$ \\
MCT (Minimum Completion Time) & $\checkmark$ & $\times$ & $\times$ & $\checkmark$ \\
Existing Auction-Based & $\checkmark$ & $\checkmark$ & $\times$ & $\checkmark$ \\
\textbf{AUC-RAC (Proposed)} & $\checkmark$ & $\checkmark$ & $\checkmark$ & $\checkmark$ \\
\hline
\end{tabular}
\label{rev2}
\end{table}

\begin{table}[h]
\centering
\caption{Illustration of performance, fairness, delay, profit, and scalability.}
\begin{tabular}{|l|l|l|l|l|l|}
\hline
\textbf{Strategy }                & \textbf{Performance }& \textbf{Fairness} & \textbf{Delay}    & \textbf{Profit}   & \textbf{Scalability} \\ \hline
\begin{tabular}[c]{@{}c@{}}Static\\ Offloading \end{tabular}        & Moderate    & Low      & High     & Low      & Poor        \\ \hline
\begin{tabular}[c]{@{}c@{}}Heuristic \\ Offloading \end{tabular}       & Good        & Moderate & Moderate & Moderate & Moderate    \\ \hline
\begin{tabular}[c]{@{}c@{}}Auction-Based \\ Offloading \end{tabular} & Excellent   & High     & Low      & High     & Excellent   \\ \hline
VM Offloading            & Poor        & Low      & High     & Poor     & Poor        \\ \hline
\end{tabular}
\label{tab3}
\end{table}

\section{Conclusion}\label{conclusion}
In conclusion, this paper presents AUC-RAC, a novel auction-based mechanism that addresses the challenges of task allocation and resource management in IoT environments. By leveraging the concepts of Docker swarm and containerization, AUC-RAC enables efficient offloading and execution of computation-intensive tasks across multiple local servers. The use of an auction-based bidding process ensures optimal task distribution and cost-effectiveness, while resource-aware containerization enhances execution efficiency. Our experimental results demonstrate the significant improvements offered by AUC-RAC in terms of offloading and service delivery for IoT devices. Furthermore, the mechanism's ability to optimize costs at both the task distribution and execution levels underscores its robustness and applicability in diverse IoT scenarios. Future work will explore further enhancements to the mechanism, including advanced strategies for handling incomplete datasets on the server and more distributed learning dynamics, to expand its utility and effectiveness in real-world deployments.

\printcredits

\bibliographystyle{model1-num-names}

\bibliography{btp}

\end{document}